\documentclass[a4paper,useAMS,usenatbib,usegraphicx]{mn2e}

\usepackage{multirow}
\usepackage{aas_macros}
\usepackage{times}
\usepackage{color}
\usepackage[total={17.8cm,24.0cm},centering]{geometry}

\newcommand{\brg}{Br$\gamma$}
\newcommand{\ha}{$\rm H\alpha$}

\newcommand{\heone}{$\rm He\,I$}
\newcommand{\fetwo}{$\rm [Fe\,II]$}
\newcommand{\htwo}{$\rm H_2$}

\newcommand{\pc}{\>{\rm pc}}

\newcommand{\mpc}{\>{\rm Mpc}}

\newcommand{\kelvin}{\>{\rm K}}

\newcommand{\mum}{\>{\mu {\rm m}}}

\newcommand{\msun}{\>{M_{\odot}}}

\newcommand{\as}{^{\prime\prime}}

\newcommand{\av}{\rm A_V}

\title[The circumnuclear environment of NGC\,613]{The circumnuclear environment of NGC\,613: a nuclear starburst caught in the act?}
\author[Falc\'on-Barroso et al.]
{J. Falc\'on-Barroso$^{1,2}$\thanks{Email: jfalcon@iac.es}, 
C. Ramos Almeida$^{1,2}$, 
T.~B\"oker$^{3}$, 
E. Schinnerer$^{4}$,
J.~H. Knapen$^{1,2}$, 
\newauthor
A. Lan\c{c}on$^{5}$, and
S. Ryder$^{6}$\\
$^{1}$Instituto de Astrof\'isica de Canarias, E-38205, La Laguna, Spain\\
$^{2}$Departamento de Astrof\'isica, Universidad de La Laguna (ULL), E-38200 La Laguna, Tenerife, Spain \\
$^{3}$European Space Agency, Keplerlaan 1, 2200 AG, Noordwijk, The Netherlands \\
$^{4}$Max-Planck-Institut f\"ur Astronomie, K\"onigstuhl 17, D-69117 Heidelberg, Germany\\
$^{5}$Observatoire Astronomique de Strasbourg, Universit\'e de Strasbourg \& CNRS (UMR 7550), Strasbourg, France\\
$^{6}$Australian Astronomical Observatory, P.O. Box 915, North Ryde, NSW 1670, Australia}

\begin{document}
\label{firstpage}

% \date{Accepted 1988 December 15. Received 1988 December 14; in original form 1988 October 11}
\pagerange{\pageref{firstpage}--\pageref{lastpage}} \pubyear{2013}

\maketitle

\begin{abstract}
We present near-infrared ($H$- and $K$-band) integral-field observations of the
inner $\sim$700\,pc of the active spiral galaxy NGC\,613, obtained with {\tt
SINFONI} on the Very Large Telescope. We use emission-line ratios to determine
the dominant excitation mechanisms in different regions within our
field-of-view, in particular the active nucleus and the star-forming
circum-nuclear ring. Diagnostic diagrams involving \fetwo\ and \htwo\
fluxes indicate that the gas is not only photoionized by the AGN in the nucleus
of NGC\,613, but also shock-heated. On the other hand, the emission line
ratios measured in the ``hot spots'' along the ring are fully consistent with
them being young star forming regions. We find no sign of radial gas transport
from the ring into the core region dominated by the AGN. The ring morphology
appears disturbed by a radial outflow of material from the AGN, which is
confirmed by the existence of a weak jet in archival radio maps. However, this
jet does not seem to have any significant effect on the morphology of the large
($\sim$8$\times$10$^7\msun$) reservoir of molecular gas that has accumulated
inside the central $\sim$100\,pc. Such a concentration of molecular gas around
an AGN is unusual, and supports a scenario in which star formation is recurrent
and episodic in spiral galaxies. In this context, NGC\,613 appears to be in
final stages of the gas accumulation phase, and is likely to undergo a nuclear
starburst in the near future.
\end{abstract}


%###############################################################################
\section{INTRODUCTION}
%###############################################################################

The question of whether and how the morphology of galaxies evolves over time is
one of the most intensely debated questions in astrophysics. Understanding this
secular evolution requires a detailed theory for the origin and fate of the 
various components found in the central regions of galaxies, such as a super-massive
black hole (SMBH), a compact nuclear star cluster, stellar bar, star-forming rings, 
and a (pseudo)bulge. Despite numerous theoretical studies 
\citep[see][and references therein]{hopkins06} and large-scale observational
programs \citep[e.g.][]{sings,dale04,sauron,dr7,atlas3d}, a complete
understanding of how these various features form, evolve, or influence each
other, remains elusive. The problem is exacerbated by the fact that some (if not
many) of these features are likely transient, and thus may no longer be obvious
in observations, even though the consequences of their past existence still are.

Circumnuclear star formation in disk galaxies is a good example of a highly
time-variable phenomenon which is both an agent for and an indicator of secular
evolution. The high gas densities required to initiate and maintain star
formation in the central few hundred pc are the result of inward radial
transport of large amounts of gas
\citep[e.g.][]{simkin80,combes85,athanassoula94,knapen95}. Over time, the newly
formed stars can alter the appearance of the galaxy, in that they contribute to
the prominence of a \hbox{(pseudo-)bulge} \citep[e.g.][]{kor04}.\looseness-2

The star formation history of the nuclear region of galaxies seems to be also
tightly linked to the properties of their central engines, i.e. active galactic
nuclei (AGN). This has been widely investigated over the last decade from a
number of observational studies which link dynamical or structural properties of
the galaxy as a whole to those of the SMBH
\citep[e.g.][]{ferrarese00,gebhardt00,graham+01,novak+06,
shapiro+06,bandara+09,kor09,atlasxxiii}. Taken together, these results make it
clear that the growth of SMBHs and the evolution of the central few $100\pc$ of
the galaxy influence each other and thus cannot be interpreted independently.

While some qualitative theoretical explanations have been proposed
\citep[e.g.][]{mcLaughlin+06}, a detailed understanding of how star formation
and AGN activity depend on each other is still lacking. This is hardly
surprising, given that both star formation and AGN activity are highly
time-dependent phenomena, and that any one galaxy can offer only a snapshot view
of this time dependence. It is therefore desirable to observe as many galactic
nuclei as possible with sufficient spatial resolution to separate the
circumnuclear star formation activity from the central engine. Such observations
are challenging because very often, the complex dust structures and resulting
high extinction values make it necessary to use infrared or radio wavelengths in
order to reveal the sites of active star formation (a.k.a. ``hot spots''), and
their properties. In addition, it is desirable to not only derive the morphology
of the various tracers of star formation, but also their kinematics in order to
get a sense for the timescales involved.

%-----------------------------------------------------------------------------
\begin{figure}
\centering
\includegraphics[angle=0,width=0.99\linewidth]{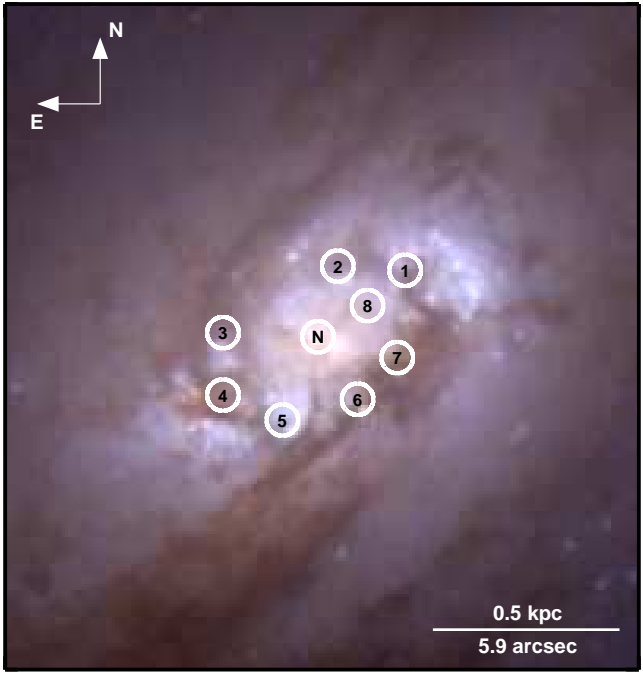}
\caption{\textit{Hubble} Space Telescope colour image of the nuclear region of
NGC\,613. Physical scale of the image is indicated in the lower right corner.
The regions under study in this work are marked with white circles.}
\label{fig:hst}
\end{figure}
%-----------------------------------------------------------------------------

Near-infrared integral field spectroscopy is the ideal tool for such studies.
When used with modern 8m-class telescopes, it offers a combination of high
sensitivity, high spatial resolution, and full spectroscopic information over
the field of view. In this paper, we present VLT/{\tt SINFONI} observations of
NGC\,613, a barred spiral galaxy with Hubble type Sbc at a distance of
$17.5\mpc$. The projected spatial scale at that
distance is 84.8 parsec per arcsec. NGC\,613 was classified as a composite
object by \citet{Veron86} based on its low-resolution optical spectrum
(Seyfert/H{\sc ii}), and it was confirmed as an AGN using mid-infrared
spectroscopy \citep{Goulding09}. NGC\,613 is part of a small sample of spirals
with kpc-scale star-forming rings discussed in \citet[hereafter paper
I]{boker08}, and some aspects of the data have already been presented there.
Here, we use the data to give a detailed account of the physical conditions of
stars and gas (both molecular and ionized) in the central $700\pc$, with the aim
of illuminating the mutual feedback between circumnuclear star formation and
nuclear activity.

The paper is structured as follows. In section~\ref{sec:data} we present a brief 
outline of the observations and instrumental setup. Section~\ref{sec:morph} 
introduces the morphology of the stellar continuum and some emission-lines, 
while \S\ref{sec:spec} describes the location of the aperture spectra 
extracted for our analysis. We focus on the properties of the stellar component 
in the nucleus and along the ring in \S\ref{sec:stellar}. The details on the 
derived extinction for each aperture are presented in \S\ref{sec:extinc}. 
We discuss the physical state of the gas in the circumnuclear region in 
\S\ref{sec:ratios} and investigate the possible connection between the nucleus 
and the star-forming ring in \S\ref{sec:discussion}. Finally, we summarise our 
results in \S\ref{sec:summary}.

%-----------------------------------------------------------------------------
\begin{figure*}
\centering
\includegraphics[angle=0,width=0.99\linewidth]{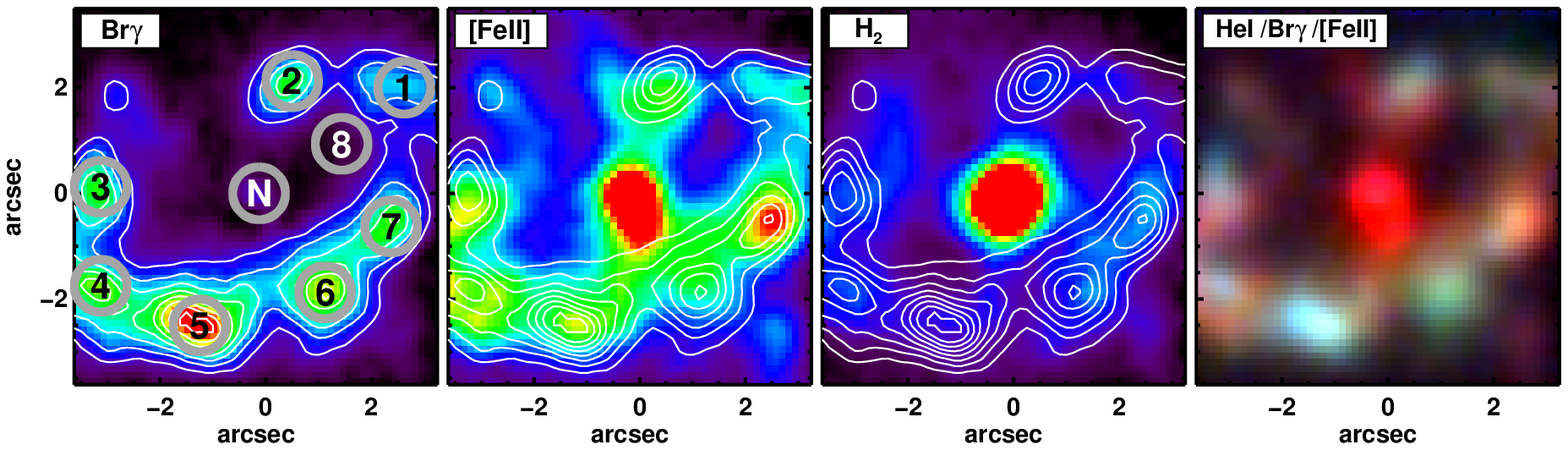}
\caption{Observed flux maps for NGC\,613 ($\sim$700\,pc wide). From left to right:, \brg\ (marking the
locations and sizes of the apertures used in this paper), \fetwo, and \htwo\ maps.
The right most panel displays a composite colour image using \fetwo\ (red),
\heone\ (blue) and \brg\ (green), see Section~\ref{sec:morph} for details.
Overlaid on all the maps are the \brg\ isophotes. In all the panels North is up
and East to the left. Displayed emission-line fluxes range between 
1.0$\times$10$^{-16}$ and 5$\times$10$^{-14}$\,erg\,cm$^{-2}$\,s$^{-1}$.}
\label{fig:maps}
\end{figure*}
%-----------------------------------------------------------------------------

%###############################################################################
\section{OBSERVATIONS \& DATA REDUCTION}\label{sec:data}
%###############################################################################

\subsection{Integral-field data}
The near-infrared data analysed in this paper are part of the data set described
in paper I. Briefly, we used the {\tt SINFONI} integral-field spectrograph at
the European Southern Observatory Very Large Telescope (Program ID: 076.B-0646A)
to observe a sample of five spiral galaxies with compact, star-forming 
circumnuclear rings selected from the imaging survey of \citet{knapen06}. In
its seeing-limited mode (i.e. without the aid of adaptive optics), {\tt SINFONI}
enables two-dimensional spectroscopy over a field-of-view of $8\as\times8\as$,
and a spatial sampling of $\rm 0.125\as /pixel$. We used the {\tt SINFONI}
configuration for simultaneous $H+K$ spectra which yields a spectral resolution
of $R\approx2000$. The total on-source integration time for NGC\,613 was 2.5
hours, divided in five identical observing blocks that were obtained over the
course of five nights in Oct./Nov. 2005. Average seeing during our observations 
was approximately 0\farcs5. The raw data were corrected for sky
background, detector dark current, and cosmic rays using standard methods. Flux
calibration was performed using standard stars obtained routinely during each
observing night. A detailed description of the data reduction and flux
calibration procedures can be found in paper I.

\subsection{Radio data}

In addition to the integral-field data, and in order to confirm the presence of
a jet (based on our \fetwo\ maps, see \S\ref{sec:morph}), we have searched the
Very Large Array (VLA) archives for radio observations. The archival C band
(4.86\,GHz) observations (project AH231) first published by \citet{hummel92}
were done in A and B configuration in 1986. We reduced the two datasets using
the standard routines in AIPS \citep{greisen90}. The quasars 3C84 and 0142-278
served as flux and phase calibrator, respectively. The combined datasets have a
resolution of 0\farcs88$\times$0\farcs46 with a PA of 0.99$^\circ$  using robust
weighting and a pixel scale of 0\farcs1 per pixel. We CLEANed the data using a
single CLEANing box and a flux limit of 1.5$\sigma$. The resulting map has an
rms of about 17 $\mu$Jy/beam.

\subsection{Hubble Space Telescope imaging data}
\label{sec:hst}

In order to get the highest spatial resolution view of the inner regions of
NGC\,613, we retrieved from the Hubble Legacy
Archive (http://hla.stsci.edu) WFPC2 data for the F450W, F606W, and
F814W filters. The dataset is part of the proposal number 9042 (PI: Stephen
Smartt) aimed at detecting the progenitors of massive, core-collapse supernovae.
Figure~\ref{fig:hst} shows a colour composite image (F450W, F606W, F814W) of the
nuclear region of NGC\,613. For reference we indicate the main apertures used
for the analysis in this paper.

%###############################################################################
\section{MORPHOLOGY OF THE CENTRAL REGION}\label{sec:morph}
%###############################################################################
In order to set the stage for the following discussion, we present in
Fig.~\ref{fig:maps} a number of maps that define the morphology in the
central $8\as$ ($\approx~700\,\pc$) of NGC\,613. They are used here 
to identify the individual "hot spots" in the NGC\,613 ring that will be
analysed in more detail using aperture spectra presented in \S\ref{sec:spec}.\looseness-2

Maps for the three emission lines \fetwo\ ($\lambda 1.64\,\mu$m), \htwo\
($\lambda 2.12\,\mu$m), and \brg\ ($\lambda 2.16\,\mu$m) have already been
presented in paper I. As described there, the line maps were generated by
summing all spectral channels over the width of the respective emission line,
and subtracting a continuum image obtained by averaging neighbouring channels on
either side. This method is equivalent to obtaining narrow band images centred
on the line and the blue and red continuum, respectively. A number of grey
circles in the \brg\ map denote $1\as$ diameter apertures centred on the various
``hot spots'' in the ring, one centred on the nucleus, and another one on a
``empty'' region.

The three emission-line maps clearly reveal a ring-like morphology composed by
seven stellar clusters. The nucleus presents high \fetwo\ and \htwo\ flux
levels, compared to \brg\ which is significantly weaker. The \fetwo\
flux distribution in the centre, however, seems somewhat more elongated than the
fairly spherical \htwo\ morphology. As already noted in paper I, there appears
to be a gap in the NW of the ring. This is conspicuous in all the three near-IR
(hereafter NIR) line maps, as well as the radio maps shown in Fig.~\ref{fig:radio}.
Interestingly, two plumes of material seem to extend from the nucleus all the
way out to the edges of the gap. We believe these features are related to the
presence of a radio outflow already noted by \citet{hummel87} and
\citet{hummel92} (see \S\ref{sec:nuc}).

The last panel in Fig.~\ref{fig:maps} displays a composite colour image produced
by combining the \heone\ (blue), \brg\ (green) and \fetwo\ (red) emission lines.
This image illustrates the {\it pearls on a string} scenario proposed in paper I
for the evolution of star formation of the hot spots in NGC\,613 ring, in which 
the hot spots age as they move along the ring away from the over-density region.

%############################################################################
\section{NEAR-INFRARED SPECTRA}\label{sec:spec}
%############################################################################

\begin{figure*}
\begin{center}
\includegraphics[angle=0,width=0.99\linewidth]{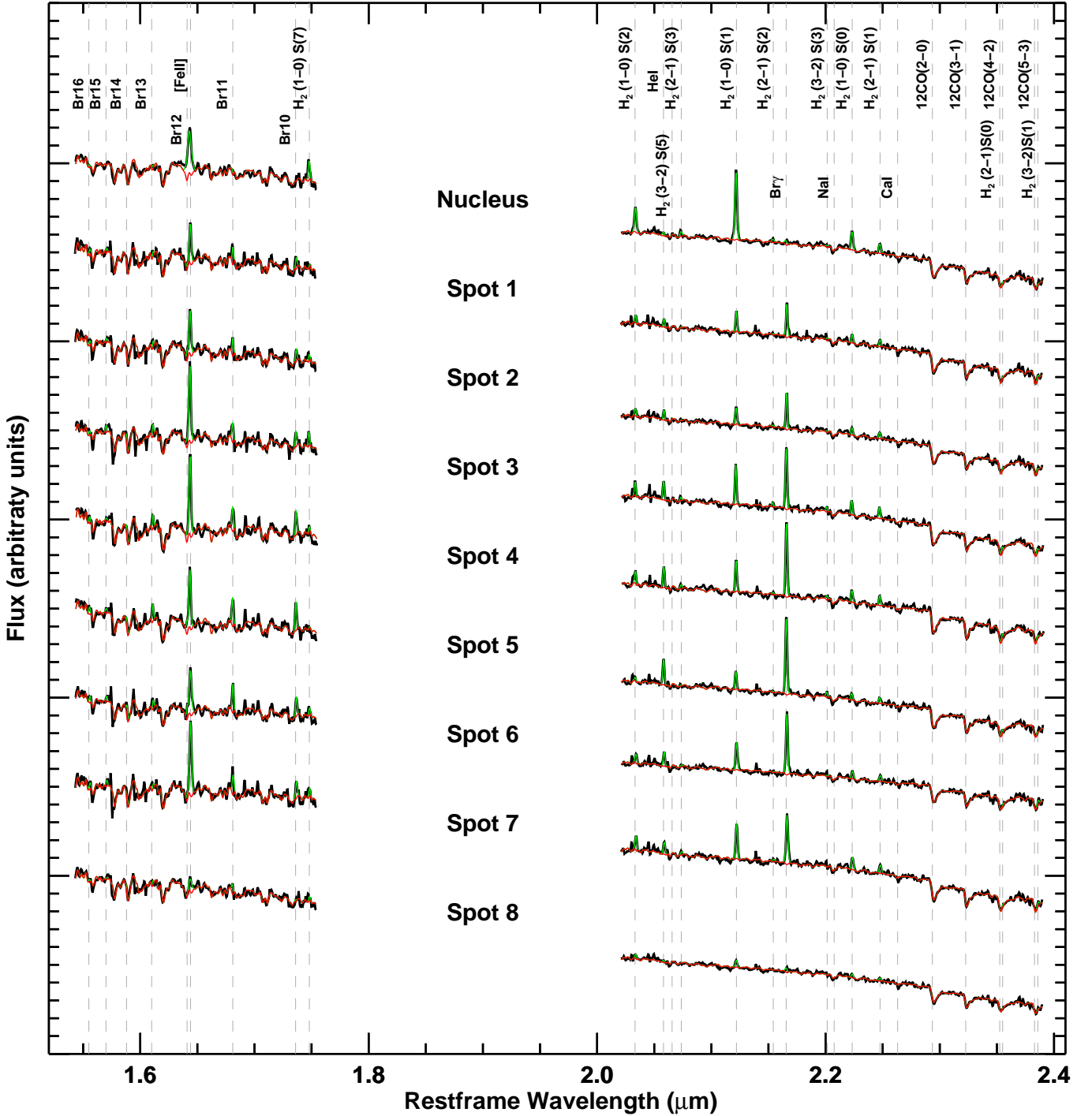}
\caption{Observed aperture spectra extracted from the regions indicated in
Fig.~\ref{fig:maps}. Black, the original spectra. Red, the best stellar fit to
the spectra (see text for details). Green the best Gaussian fit to the emission
lines present in the spectra. Main spectral features are also indicated.}
\label{fig:spectra}
\end{center}
\end{figure*}

In order to perform a more detailed analysis of the central region of NGC\,613,
we have extracted $H$- and $K$-band spectra for the apertures indicated in 
Fig.~\ref{fig:hst} and the \brg\ map in Fig.~\ref{fig:maps}. The spectra are 
presented in Fig.~\ref{fig:spectra}. Their flux calibration is accurate to within
15\% as discussed in paper I.

We have performed a quantitative analysis of the emission line fluxes and
kinematics, and made an effort to decompose the spectra into the stellar
continuum and the ``pure'' emission line spectrum. For this, we made use of the
{\sc GANDALF} package \citep{sarzi06}. The software performs a simultaneous
least-squares fit to both the stellar continuum and emission lines. The stellar
continuum is described using theoretical spectra of red giants and supergiants
which dominate the NIR emission of evolved stellar populations and star-forming
regions, respectively. We made use of the library of theoretical model spectra
by \citet{lancon07}. The spectra used for our fits have solar abundances, and
cover a range of effective temperatures (T$_{\rm eff}$=2900--5900\,K) and
gravity (log(g)=0--2). The models were re-binned to match the resolution of the
data. The emission lines are treated simply as Gaussian templates. Their fluxes
were left unconstrained during the fitting process, so that the amount of
extinction could be estimated. The exact peak position and the width of
the \fetwo\ line were determined independently from those of
the Hydrogen Brackett-series and the different transitions of the H$_2$
molecule. For the latter, however, the different lines in each series
were forced to share the same kinematics, i.e., line-of-sight velocity and
velocity dispersion, as they did not show any substantial differences. Given the
instrumental resolution of our data ($R\approx2000$), most of the emission lines
are unresolved, with the exception of the \fetwo\ line. Uncertainties in
the different line fluxes have been determined by generating 100 Monte Carlo
realisations of the input spectra. We achieved this by perturbing our input
spectra with white noise using the amplitude of the residuals from our spectral
fits with GANDALF as an estimate of their variance.

Tables~\ref{tab:fluxes} and~\ref{tab:h2fluxes} summarize the results of the
emission-line analysis for all nine apertures. We list all lines that are
detected with a minimum signal-to-noise ratio of 3. 

%-----------------------------------------------------------------------------
%-----------------------------------------------------------------------------

%###############################################################################
\section{THE STELLAR COMPONENT}\label{sec:stellar}
%###############################################################################

To our knowledge the only work measuring CO line-strengths in detail in NGC\,613 
is that of \citet{js99}. In their study they placed several slits across
different locations of the galaxy: the nuclear region and one of the spiral
arms. One of the main findings in their paper was the fact that in the spiral
arm the measured CO line-strength (using the \citealt{doyon94} definition of the
index) was significantly lower than that expected from old stellar populations.
Ruling out metallicity and extremely young stellar populations as the causes for
this CO depletion, dilution by host dust was their main argument to explain the
observed CO values. The results on the circumnuclear region presented in their
paper confirm the presence of a young population (up to 24\% of the light in the
$K$-band) in the innermost 4$\as$, a fraction that decreases significantly as we
move into the bulge dominated region (which is consistent with an old stellar
population).

%-----------------------------------------------------------------------------
\begin{figure}
\begin{center}
\includegraphics[angle=0,width=\linewidth]{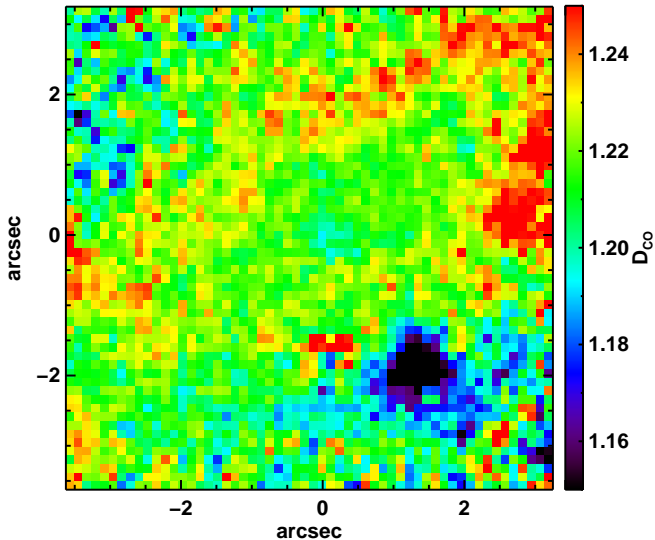}
\caption{Two dimensional distribution of the D$_{\rm CO}$ line index \citep{marmol08} 
in the centre of NGC\,613}
\label{fig:DCOmap}
\end{center}
\end{figure}
%-----------------------------------------------------------------------------

Here we make use of the CO line-strength index (D$_{\rm CO}$) developed by
\citet{marmol08} to interpret our observations. This definition is better suited
for the study of stellar populations than previous ones, as it is based on
observations of a new intermediate resolution $K$-band stellar library that
exceeds the coverage of the stellar atmospheric parameters of previous works. It
also has the advantage of being fairly independent on the broadening of the CO
line, and thus avoiding often uncertain corrections. The index was measured over
the aperture spectra using the {\hbox {\small INDEXF}} code
({http://www.ucm.es/info/Astrof/software/indexf/}) by \citet{cardiel07}.\looseness-1

In Fig.~\ref{fig:DCOmap} we present the two dimensional distribution of the
D$_{\rm CO}$ index in the centre of NGC\,613. The map reveals some level of
structure. The North side of the ring has slightly enhanced CO absorption
compared to the southern part, where the most star-forming hot spots are
located. Spot~6 displays the lowest D$_{\rm CO}$ value, as also evident from its
spectrum in Fig.~\ref{fig:spectra}. Spot~6 is also the most heavily obscured
region observed in the \textit{HST} image presented in Fig.~\ref{fig:hst}. The
nucleus also displays a slightly lower D$_{\rm CO}$ content compared to its
surroundings.\looseness-1

\begin{figure}
\begin{center}
\includegraphics[angle=0,width=0.99\linewidth]{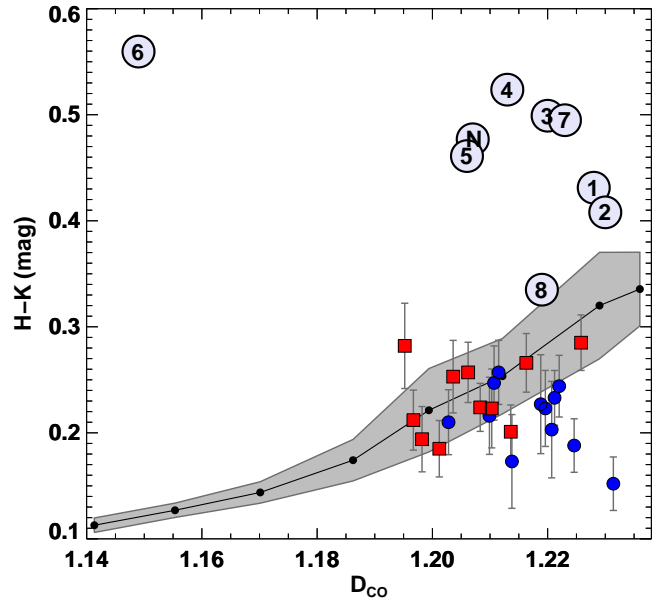}
\caption{$H-K$ colour vs D$_{\rm CO}$ index for the aperture spectra in
Fig.~\ref{fig:maps} (filled circles). Shaded grey region marks the colour and
D$_{\rm CO}$ measurement for the stellar templates of \citet{lancon07}. $H-K$
colours have been computed from the spectra themselves. For comparison, red and blue
squares are measurements for the sample of (cluster and field, respectively)
early-type galaxies in \citet{mm09}.}
\label{fig:HK_CO}
\end{center}
\end{figure}

Figure~\ref{fig:HK_CO} shows the D$_{\rm CO}$ measurements for the different
apertures defined in Fig.~\ref{fig:maps}. Given the high signal-to-noise of our
aperture spectra, uncertainties in the D$_{\rm CO}$ index are negligible and
well below 0.01. Except for Spot~6 all the other regions have D$_{\rm CO}$ values
similar to those observed in the sample of field and cluster early-type galaxies
of \citet{mm09}. This result may be somewhat unexpected, at least for the
star-forming regions, given the marked difference between their integrated
stellar populations and those of early-type galaxies. As illustrated in
\citet{mayya97} as well as by the Starburst99 predictions \citep{starburst99} the
evolution with time of the CO equivalent width for instantaneous bursts of star
formation exhibits several peaks at different phases (most prominently the
red supergiant phase around 10\,Myr). This behaviour makes it possible for such
extremely different stellar populations to display similar D$_{\rm CO}$ values.
At the same time it highlights the complexity of modelling the stellar content
of galaxies in this wavelength range alone. 

On a side note, interestingly, Fig.~\ref{fig:HK_CO} shows a fairly marked
difference, well beyond the reported measurement errors, in the D$_{\rm CO}$
content of field and cluster early-type galaxies. \citet{mm09} interpret this
difference as a signature of an excess of intermediate age, asymptotic giant
branch (AGB) stars in the stellar light of field galaxies. An alternative
explanation, also considered by these authors, is the environmental dependence
of the carbon abundance. The similarity in the D$_{\rm CO}$ values between the
most actively star-forming regions in NGC\,613 and those of field galaxies, with
significantly older stellar populations, strongly supports the latter view that
environment rather than AGB stars, plays an important role in the chemical
enrichment of early-type galaxies \citep[see also][]{carretero04}.

One of the discoveries in paper I was the age sequence (see the right-most panel
of Fig.~\ref{fig:maps}), formed by Spots~5, 6 and 7, along the southern side of
the ring. As shown in this section, the complexity in understanding stellar
populations at near-IR wavelengths makes it difficult to check whether the age
gradient inferred from emission line ratios in paper I can be confirmed by the
mean stellar ages of the hot spots. Nevertheless we refer the interested reader
to \citet{allard06}, \citet{sarzi07} and \citet{vdl13} for examples of this kind
of study in nearby galaxies with star-forming nuclear rings.

%-----------------------------------------------------------------------------
%-----------------------------------------------------------------------------

%###############################################################################
\section{EXTINCTION CORRECTION}\label{sec:extinc}
%###############################################################################
In this section we employ several methods to determine the amount of extinction
in the inner regions of NGC\,613.  The list of emission-line fluxes for the 
different apertures, corrected for extinction, along with the computed A$_{\rm V}$ 
values, are presented in Tables~\ref{tab:fluxes} and \ref{tab:h2fluxes}.

\subsection{The Brackett decrement: \brg/Br10}
\label{sec:Br_ratios}
The flux ratio $\psi$\,=\,\brg /Br10 can be used to estimate the extinction within
the {\tt SINFONI} field of view, because the relative intensities of the
hydrogen recombination lines for various gas densities and temperatures are well
understood (e.g. \citealt{hs87}). Any deviation from the theoretical line ratio
$\psi_{0} = 3.07$ for a ``standard'' environment (case B,
$\rho  = 10^4\,e^-\,\rm{cm}^{-3}$, and $T_e  = 7500\kelvin$) can then be used to
infer the amount of visual extinction, using the \citet{rl85} extinction law, along
the line-of-sight as:\looseness-2

\begin{equation}
\av = 50.5 \cdot log(\psi/\psi_{0})
\end{equation}

The measurement of A$_{\rm V}$ using this method is only possible along the ring
(i.e. Spots~1 to 7). Outside this area (Spot~8 and the nucleus), neither \brg\
nor Br10 can be measured accurately enough to determine a reliable value.
Extinction in the latter apertures was determined using the method outlined in
the following section.

\subsection{$H-K$ vs CO index}
\label{sec:HK_CO}

While the use of line ratios is a relatively straightforward method to estimate
extinction, it is highly sensitive to the signal-to-noise ratio of the data. An
alternative approach for those cases where the line ratios are not accessible is
the use of colours and line-strength indices to obtain extinction values. The
combination of a colour image, which is affected by dust, with a line index,
largely insensitive to extinction \citep{macArthur05} is a powerful method that
has not been applied much in the literature, because of the lack of
well-calibrated line strength data \citep[see][]{ganda09}. By choosing the
appropriate colour, this technique gives the colour excess in an almost
model-independent way, which can easily be converted to A$_{\rm V}$.\looseness-2

In our case, we can measure the $H-K$ colour by integrating the flux in the
spectra of each aperture within the given bandpass. From our $H-K$
colour map alone it is very difficult to determine the amount of extinction as
it is not obvious how much of the colour is due to intrinsic stellar populations
and how much comes from dust. Having two colours does not help a lot, since the
effects of reddening in colour-colour diagrams is almost parallel to the effect
of changing metallicity or age \citep[e.g.][]{kuchinski98}.The most
prominent absorption features in the $H$- and $K$-band spectra presented in
Fig.~\ref{fig:spectra} are the $^{12,13}$CO bandheads. Stellar population
modelling has shown that the strength of these features reaches a maximum when
the spectrum is dominated by red supergiants, typically 15 to 40\,Myr after the
birth of the population \citep{persson83,doyon94,rhoads98}. Several authors defined 
quantitative diagnostics in the NIR spectrum of evolved stars to
discriminate between the different luminosity classes
\citep[e.g.][]{baldwin73,frogel78,kh86,omo93,doyon94,puxley97,fs00,ryder01,riffel07}.

Figure~\ref{fig:HK_CO} illustrates the method to estimate the extinction. We
assume that in the absence of dust the regions should have similar colours
to those predicted by the stellar templates of \citet[][shaded area in
Fig.~\ref{fig:HK_CO}]{lancon07}. The colour excess E($H-K$), the difference
between data and the stellar spectra, is used in conjunction with the
\citet{rl85} extinction law to compute A$_{\rm V}$ values. Since both the colour
and D$_{\rm CO}$ index are fairly insensitive to instrumental broadening, there
is no need to degrade our data's spectral resolution to match that of the
theoretical stellar spectra. For reference we draw in the same figure the sample
of field and cluster early-type galaxies of \citet{mm09}, which are expected to
have very small colour excess. 

In general the measurements obtained with the line ratios in
\S\ref{sec:Br_ratios} are a factor 2--5 larger than those derived
here\footnote{Typical formal uncertainties in the extinction measured from the
\brg/Br10 ratio are around 2\,mag whereas those associated with the method
presented in this section are on the order of 0.5\,mag.}.  The observed
differences could be associated to the assumptions made by each of the methods,
i.e. the colour-based extinction relies on theoretical stars that might not
mimic the intrinsically dusty conditions of star-forming regions
\cite[e.g.][]{calzetti94}. This behaviour is actually not unusual and it is
often found in star-forming galaxies \citep[e.g.][]{hao11,kreckel13}. Following
\citet{kreckel13}, who concluded that the line ratio decrement method provides a more
reliable extinction measurement for dusty regions, we use the line ratio based A$_{\rm V}$ 
values to correct our measured fluxes in the hot spots, and the
colour-based method presented in this section for Spot~8 and the nucleus only.\looseness-1

%###############################################################################
\section{EMISSION-LINE ANALYSIS}\label{sec:ratios}
%###############################################################################

Emission-line ratios of different elements constitute a powerful tool to
disentangle the dominant excitation mechanisms in different regions in galaxies.
Here we use the ratios of the \fetwo, \htwo\ and \brg\ lines for this purpose. 
At the same time we make use of the \brg\ line to provide star formation rate
estimates, as well as the amount of H{\sc ii} present in the star-forming ring.

\subsection{Star formation rates}\label{sec:sfr}

Now that we have an estimate for the extinction in the different regions
along the ring, we can correct the observed flux in the \brg\ line and thus
estimate the intrinsic flux. Knowing the intrinsic ratio between the \brg\ and
\ha\ lines for case B recombination (\ha/\brg$\approx$104, \citealt{hs87}),
we can convert those luminosities into star formation rates (SFRs) following the
prescription by \citet{kennicutt98}:

\begin{equation}
{\rm SFR}(M_{\odot}~{\rm yr}^{-1}) = \frac{1}{1.26\times10^{41}~{\rm erg~s^{-1}}}\cdot L({\rm H\alpha}).
\end{equation}

\noindent In addition, we derived the mass of the ionised gas in each aperture assuming
the same case B recombination scenario as above.\looseness-2

The star formation rates measured in the ring spots range from 0.03 to 0.1\,$\msun$/yr. 
The value for the nucleus, however, drops down to 0.015\,$\msun$/yr,
which is lower than the SFRs reported by \citet{Kewley02} for a sample of 81
nearby non-active galaxies. It is therefore likely that the recombination lines
observed in the nucleus are not due to star formation but the AGN. The SFRs
measured in apertures 5, 6 and 7 show a steady decline along the ring, which is consistent with
the {\it pearls on a string} scenario presented in paper I (i.e. the hot spots
passively evolve as they move along the ring and away from the over-density
regions).

The masses of the ionised-gas in the H{\sc ii} regions probed by the apertures
vary from $\approx$\,1.2\,$\times$\,10$^{3}\msun$ in the nucleus to
11.2\,$\times$\,10$^{3}\msun$ for the most massive hot spot. The SFR densities
in the hot spots are in good agreement with those of circumnuclear star-forming
regions in disk galaxies \citep{kennicutt98}.\looseness-2

\subsection{The \fetwo\ \& \brg\ ratio}

Strong \fetwo\ emission is indicative of shock-excited gas in the filaments of
supernova remnants, in contrast to the weak \fetwo\ emission characteristic of
photoionized gas in H{\sc ii} regions. In AGN, strong \fetwo\ emission is also
common, although several processes may contribute to its production: (1)
photoionization by extreme UV to soft X-ray radiation from the central source,
producing large partially ionized regions in the NLR clouds; (2) interaction of
radio jets with the surrounding medium, inducing shocks and hence partially
ionized cooling tails; and (3) fast shocks associated with supernova remnants
present in starburst regions. The \fetwo/\brg\ ratio has proved to be very
useful for distinguishing between a stellar or non-stellar origin of the \fetwo\
emission. This line ratio increases from H{\sc ii} regions (photoionization by
hot stars) to supernova remnants (shock excitation), passing through starburst
and active galaxies \citep[e.g.][]{alonso97,ra04,ra05,ramos06,Ramos09,Riffel13}. 

\begin{figure}
\begin{center}
\includegraphics[angle=0,width=0.99\linewidth]{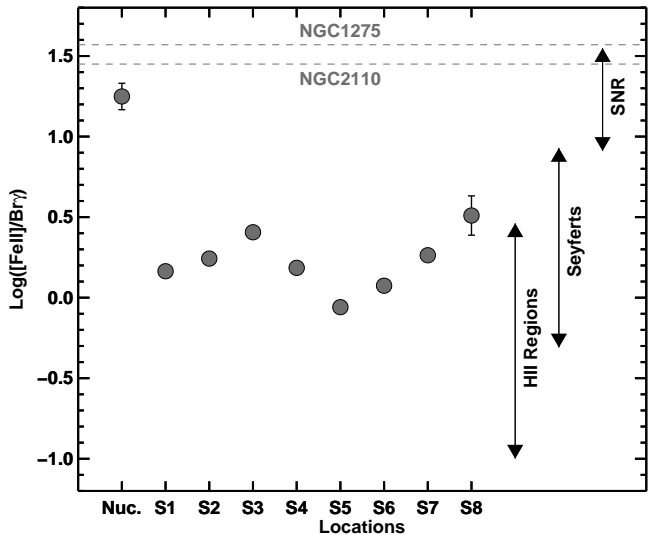}
\caption{Variation of the \fetwo\ over \brg\ ratio for the different apertures
considered here. We indicate typical ratios for different types of astrophysical
objects from the compilations of \citet{mo88,mkt93,dale04,ra04,ra05}. Horizontal
dashed lines mark the values observed for the two radio-loud galaxies NGC\,1275
and NGC\,2110 from \citet{Kawara93} and \citet{ra05} respectively.}
\label{fig:BrgFeIIratio}
\end{center}
\end{figure}

Figure~\ref{fig:BrgFeIIratio} shows the variation of the \fetwo/\brg\ ratio for
our apertures. The ratio is relatively low (\fetwo/\brg=[0.9,2.5]) in the
star-forming hot spots, consistent with the typical values of H{\sc ii} regions.
On the other hand, the ratio in the nucleus is high (\fetwo/\brg=17.7), in the
domain populated by supernova remnants and radio-loud Seyferts as e.g. NGC\,1275
and NGC\,2110 (\fetwo/\brg=37.3 and 28.1 from \citealt{Kawara93} and
\citealt{ra05} respectively). As found by \citet{Forbes93}, there is a tight
correlation between \fetwo\ and radio emission in both Seyfert and star forming
galaxies. In fact, as we can see in Sec.~\ref{sec:nuc}, the radio contours
perfectly trace the \fetwo\ emission. We thus conclude from 
Fig.~\ref{fig:BrgFeIIratio} that the excitation mechanisms in the nucleus of
NGC\,613 (i.e. photoionization and shocks) differ from that in the hot spots
(stellar photoionization).\looseness-2

%-----------------------------------------------------------------------------
%-----------------------------------------------------------------------------

\subsection{Molecular hydrogen}

All aperture spectra shown in Fig.~\ref{fig:spectra} show a number of emission
lines from molecular hydrogen. This section investigates what can be learned
about the physical state of the \htwo\ gas. In this effort, we follow a number
of studies, both theoretical \citep[e.g.][]{mouri94}, and observational
\citep[e.g.][]{vei97,krabbe00,davies05,ra04,ra05,ramos06,Ramos09,Riffel13,
Mazzalay13}. All these studies take advantage of the fact that in most
astronomical environments, the lowest vibrational levels ($v=1$) of \htwo\ tend
to be well thermalised, while higher level transitions are predominantly
populated by processes such as non-thermal UV fluorescence.

Generally speaking, the \htwo\ molecule can be excited via three distinct
mechanisms: (i) UV fluorescence, where photons with $\lambda> 912$\AA\ are
absorbed by the \htwo\ molecule and then re-emitted \citep[e.g.][]{bv87}; (ii)
shocks, where high-velocity gas motions heat and accelerate the ambient gas
\citep[e.g.][]{holkee89}; and (iii) X-ray illumination, where hard X-ray photons
penetrate deep and heat large amounts of molecular gas 
\citep[e.g.][]{maloney96}. Shocks and X-ray illumination are normally referred
to as thermal processes, and UV fluorescence as non-thermal. Each of these three
mechanisms produces a distinct \htwo\ spectrum.\looseness-2

The 1$-$0S(1)/2$-$1S(1) line ratio is an excellent discriminator between
thermal and non-thermal processes. According to the models of \cite{mouri94},
this ratio has much lower values ($\leq2$) in regions that are dominated by
UV fluorescence than in thermal-dominated gas ($\geq5$). This diagnostic has
the advantages of being fairly insensitive to extinction because both lines have
similar wavelengths and are independent of the ortho/para ratio. At the same time,
the 1$-$0S(2)/1$-$0S(0) line ratio is sensitive to the strength of the
incident radiation, and it can be used to discriminate the dominant excitation
process.

%-----------------------------------------------------------------------------
\begin{figure}
\begin{center}
\includegraphics[angle=0,width=0.99\linewidth]{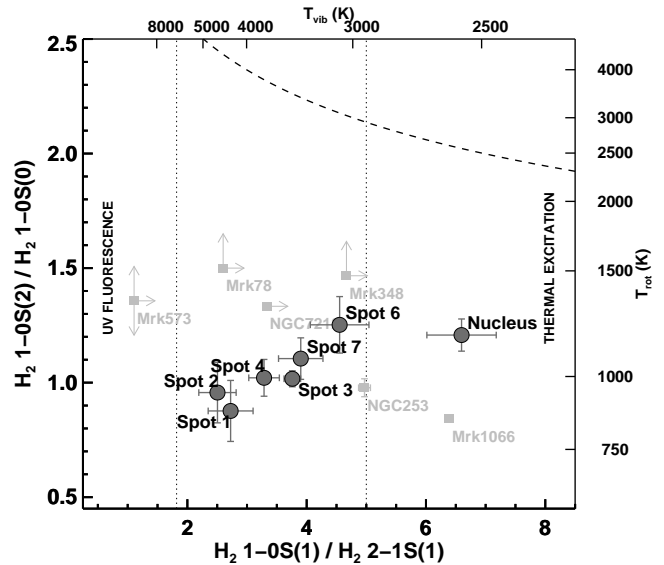}
\caption{\htwo\ 1$-$0S(2)/1$-$0S(0) versus \htwo\ 1$-$0S(1)/2$-$1S(1)
line ratios of the different selected regions in NGC\,613. In addition, 
literature values for other galaxies are presented (see text for details). The
dashed line indicates the locus of equal T$_{\rm vib}$ and T$_{\rm rot}$.
Vertical dotted lines delimit the regions of thermal and UV fluorescence
excitation from \citet{mouri94}.}
\label{fig:H2ratio}
\end{center}
\end{figure}
%-----------------------------------------------------------------------------

Figure~\ref{fig:H2ratio} shows - for the apertures defined in
Fig.~\ref{fig:maps} - the value of the flux ratio of the 1$-$0S(1) and 
2$-$1S(1) versus the ratio of the 1$-$0S(2) and 1$-$0S(0) lines. For
comparison, we also plot the ratios of five nearby Seyfert 2 galaxies from 
\citet{Ramos09} and of the starburst galaxy NGC\,253 from \citet{Riffel13}.\looseness-2 

All the hot spots in the ring are, to varying degrees, affected
by a mixture of thermal and non-thermal (i.e. fluorescent) excitation. The 
molecular gas in the nucleus of NGC\,613, however, falls in the thermal excitation domain, indicating a lack of strong non-thermal
excitation mechanisms. The \htwo\ ratios measured for the nucleus are also far
from those predicted by thermal X-ray models (e.g. \citealt{Lepp83,maloney96}),
which would lie outside the upper right boundaries of Fig.~\ref{fig:H2ratio}.

To confirm these results, we can compare the rotational and vibrational
temperatures (T$_{\rm vib}$ and T$_{\rm rot}$) of the \htwo\ gas, which we
calculated using the expressions in \citet{Reunanen02}. In the case of thermal
excitation, both temperatures should be similar (i.e. close to the dashed line
in Fig.~\ref{fig:H2ratio}), whereas in the case of non-thermal excitation, 
T$_{\rm vib}>>T_{rot}$. The latter is the case of the hot spots,
which have T$_{\rm vib}$=3000--5000 K and T$_{rot}<$1200 K. On the other hand,
the \htwo\ gas in the nucleus has T$_{\rm vib}\sim$2500$\pm$500 K and T$_{\rm
rot}\sim$1200$\pm$40 K, characteristic of thermally excited gas. 

At face value, these results are consistent with the absence of strong
star formation and of a luminous X-ray source in the nucleus. Its relatively high
1$-$0S(2)/1$-$0S(0) value falls in the region of shock heated (thermalised) gas,
and points to mechanical energy from the radio outflow as the source of \htwo\
heating. All seven ring apertures, in contrast, show a much more pronounced
contribution from UV fluorescence. This is fully consistent with them being
young star-forming regions.

In addition to elucidating the excitation mechanisms at play in the inner regions of 
NGC\,613, it is interesting to estimate the amount of molecular gas present. 
This information is very valuable as it can be used to estimate the level of star 
formation expected in the near future in these regions. Using the $1-0$S(1) 
line fluxes reported in Table~\ref{tab:h2fluxes}, we 
estimate the mass of hot \htwo\ following the equation

%-----------------------------------------------------------------------------
\begin{equation}
m_{\rm H_2}^{hot}(M_{\odot}) \simeq 5.0875\times10^{13}d^2I_{1-0{\rm S}(1)},
\end{equation}
%-----------------------------------------------------------------------------

\noindent from \citet{Reunanen02}, as well as the assumptions for gas
temperature, transition probability and population fractions therein.
$I_{1-0S(1)}$ is the extinction-corrected line flux, and $d$ the distance to the
galaxy in Mpc. The resulting \htwo\ masses are reported in
Table~\ref{tab:h2fluxes}. We derive values between 9.4 and 38.5\,$\msun$ for the
ring spots and 112\,$\msun$ for the nucleus. In the ring spots, where star
formation is taking place, the mass of hot gas is low, whereas in the nucleus
the mass of hot \htwo\ is an order of magnitude larger, although this
measurement is likely affected by the interaction with the radio jet.
Nevertheless, our masses of hot \htwo\ are consistent with those derived for a
sample of Seyferts and low-luminosity AGN using {\tt SINFONI} data as those used
here \citep{Mazzalay13}. 

The same authors also estimated the the mass of cold \htwo\ from the 
integrated 1$-$0S(1) luminosity ($L_{1-0{\rm S}(1)}$) in erg~s$^{-1}\mum^{-1}$ 
using the relation:

%-----------------------------------------------------------------------------
\begin{equation}
\label{eq:cold_mass}
m_{\rm H_2}^{cold}(M_{\odot}) \approx 1174\times L_{1-0{\rm S}(1)}(L_{\odot}), 
\end{equation}
%-----------------------------------------------------------------------------

derived by comparing a large number of $m_{\rm H_2}^{cold}$ values from the
literature, calculated from CO observations, and integrated \htwo\ 1$-$0S(1)
luminosities\footnote{The reported uncertainty of this calibration is about a
factor 2 in mass.}. Thus, we can roughly estimate the masses of cold \htwo\ gas
for the nucleus and the hot spots in NGC\,613 (also listed in
Table~\ref{tab:h2fluxes}). Those masses range between $\sim$7 and
28$\times10^6\msun$ in the ring spots, and $\sim$8$\times10^7\msun$ in the
nucleus. 

An important word of caution is necessary on the interpretation of \htwo\ masses
having an associated amount of cold molecular gas. While relations like the one
presented in Eq.~\ref{eq:cold_mass} are very useful as proxies for the presence
of cold gas, there are known cases of galaxies with estimates of molecular mass
content, based on the \htwo\ line, two orders of magnitude above the one 
provided by the direct measurement of CO emission \cite[e.g.~NGC\,4151,][]{dumas10}.

%##########################################################################
\section{DISCUSSION}\label{sec:discussion}
%##########################################################################

In light of the analysis and results described in the previous sections, we
now discuss some of the most interesting unsolved questions 
in the circumnuclear environment of NGC\,613. These are mainly related to the
interactions between the nuclear ring and the radio outflow, and to the source
of excitation of the \htwo\ present in the nucleus.

\subsection{The radio outflow: an active nucleus in NGC\,613?}
\label{sec:nuc}

The classification of NGC\,613 in terms of nuclear activity is not completely
clear by looking at its optical spectrum alone. \citet{Veron86}, as part of a
spectroscopic study of the complete Revised Shapley-Ames Catalogue, classified
it as composite object (Seyfert-like component co-existing with an H{\sc ii}
region). This classification is not surprising considering that the spectrum was
extracted with an aperture of 4\arcsec, which corresponds to 328\,pc in the case
of NGC\,613. The AGN nature of NGC\,613 has, however, been recently confirmed
from mid-infrared spectroscopy. \citet{Goulding09} reported Spitzer/IRS
observations showing high-excitation lines such as [Ne{\sc III}], [Ne{\sc V}]
and [Si{\sc II}], characteristic of active galaxies. These authors claimed that
the optical signatures of nuclear activity in this galaxy are likely diluted by
strong star formation. In the X-rays, the galaxy was observed with XMM-Newton on
December 2010 and, although the data have not been published yet, they appear to
confirm the presence of an active nucleus in NGC\,613 (obsID 0654800501).
 
\begin{figure}
\begin{center}
\includegraphics[angle=0,width=\linewidth]{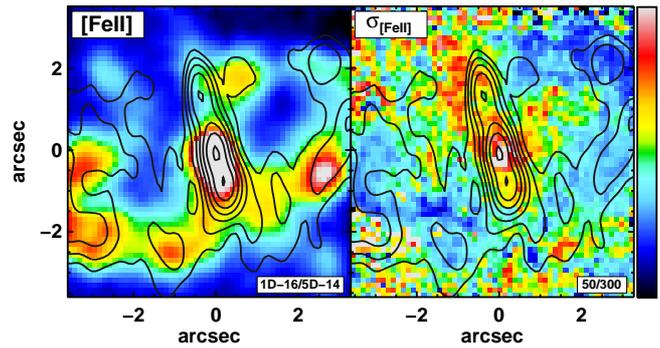}
\caption{{\tt SINFONI} \fetwo\ flux and velocity dispersion maps for NGC\,613. 
Overlaid we show VLA radio contours (see \S\ref{sec:data} for details). The 
inset in the lower-right of each panel gives the range of values for the colour 
scale.}
\label{fig:radio}
\end{center}
\end{figure}

The high resolution radio continuum images presented in \citet{hummel92}, and
re-analysed here (see \S\ref{sec:data}), show evidence for an energetic outflow
emanating from the nucleus. The radio map (see Fig.~\ref{fig:radio}) reveals a
linear feature of about 300\,pc which consists of three discrete components.
These radio blobs are perpendicularly oriented to the projected major axis of
the star-forming ring observed in our {\tt SINFONI} maps. Whether the radio
blobs are truly separate entities aligned in one direction, or bubbles of hot
plasma originating from the incident radiation has not yet been determined.
However, both the radio morphology and the coincidence of the central blob with
the optical position (radio-optical offset of 0\farcs1) indicate that the
central component of the linear feature is indeed the nucleus. This constitutes
a first indication that the radio jet orientation might be close to the plane of
the sky, and consequently, relatively close to the plane of the galaxy as well,
which has an inclination angle of $\sim$35$^\circ$ (as listed in
Hyperleda\footnote{http://leda.univ-lyon1.fr/}). Interestingly, \citet{laine06}
found a tendency towards perpendicular alignments in Seyferts, but more 
parallel in starbursts. 

\citet{Kondratko06} and \citet{Castangia08} reported the detection of a H$_2$O
megamaser with an isotropic luminosity of $\sim$35\,L$_{\odot}$, coincident with
the position of the nucleus in the optical (with an uncertainty of 1.3\arcsec).
This coincidence supports the link between the AGN and maser emission. In fact,
all masers with isotropic luminosities $>$10\,L$_{\odot}$ are associated with
AGN \citep{Zhang12}, as confirmed from interferometry. There is good evidence
that extragalactic H$_2$O megamasers trace edge-on accretion disks in Seyfert
galaxies \citep[e.g.][]{Greenhill97,Greenhill03}, since large line-of-sight
column densities are required theoretically to make the masers observable
\citep{Lo05,Henkel05}. However, there is a different class of H$_2$O
megamasers, the so-called ``jet-masers''. In these sources, the maser emission
is the result of an interaction between the radio jet and a molecular cloud on
parsec scales \citep{Braatz97,Henkel98,Peck03}. Only four Seyfert galaxies have
jet-masers confirmed to date: Mrk\,348, NGC\,1052, NGC\,1068 and the Circinus
galaxy\footnote{NGC\,1068 and Circinus seem to have both disk- and jet-masers.}
(see \citealt{Peck03} and references therein). Although it is difficult to infer
the radio jet orientation from maser emission, in all the known jet-masers the
radio jets are perpendicular to the line-of-sight. 

In the case of NGC\,613, the origin of the megamaser is uncertain.
\citet{Kondratko06} reported the detection of a very broad H$_2$O emission
feature (FWHM$\sim$87 km~s$^{-1}$), which was subsequently resolved into two
different components of 20 and 40 km~s$^{-1}$ by \citet{Castangia08}. These two
components are redshifted and blueshifted, respectively, with respect to the
galaxy systemic velocity. The large linewidths and the presence of two kinematic
components only point to a jet-maser, but the fact that the H$_2$O emission is
centred on the systemic velocity is not typical of those. Independently of the
disk- and/or jet-maser classification of the H$_2$O megamaser in NGC\,613, it is
very likely that the orientation of the jet is close to the plane of the sky.
This orientation favours shocks capable of yielding a strong jet-maser along our
line-of-sight (see Fig.~11 in \citealt{Peck03}). If we have a disk-maser, the
accretion disk and the dusty torus have to be very close to edge-on for the
line-of-sight column density to make the maser detectable, and thus, the radio
jet would also be orthogonally oriented to the plane of the disk (see
\citealt{Drouart12} and references therein).\looseness-1 

Once we know the radio jet orientation, our NIR observations provide additional
information to establish the true morphology of the outflow. In paper I, we
showed how the structure of the different emission lines in our spectral range
were affected by the out-flowing radiation. The flux maps for all the lines
(i.e. \fetwo, \heone, \htwo, and \brg) display a lack of emission on the top
half of the ring, coinciding with the direction of the out flowing material.
This indicates that the interaction with the radio jet has probably swept out
gas and dust. This is visible in Fig.~\ref{fig:hst}.

%-----------------------------------------------------------------------------
%-----------------------------------------------------------------------------

While morphologically all the lines seem altered, only the \fetwo\ line
appears to be affected dynamically. In Fig.~\ref{fig:radio} we show the \fetwo\ line flux
and velocity dispersion with radio flux contours overlaid. We see that the
highest velocity dispersion values coincide, although not perfectly, with the
linear structure resolved in three blobs observed in the radio continuum map. 
Although marginal, the values happen to be somewhat lower at the edges of this
gap, coinciding with two plumes of \fetwo\ that extend beyond the nuclear ring.
Taken at face value, this picture is consistent with a jet radiating along a
cone. The high velocity dispersions can be explained by the fact that the
differences in the line-of-sight velocities of the material are maximum in the
central parts of the cone. At the edges of the cone, velocities are very
similar, which explains the low values there. Regarding the flux, the
enhancement we see at the edges would be caused by limb-brightening of the
conical structure.

In conclusion, our NIR data support the presence of a radio jet oriented close to
the plane of the sky, which is sweeping out the gas and dust within the top half
of the star-forming ring. According to \citet{hummel92}, if the ring is
circular, its inclination is $\sim$55$^\circ$, larger than the 35$^\circ$
inclination of the galaxy. This larger disk inclination explains the proposed
scenario: the interaction with the jet affects the material in the top half of
the ring, while the bottom part remains unaltered.

\subsection{The nucleus: a starburst in waiting?}

It is evident from the numerous emission lines in the spectrum of the central
$1\as$ ($84.8\,\pc$) in Fig.~\ref{fig:spectra} that the nucleus of NGC\,613
harbours a substantial amount of hot \htwo\ gas. Our mass estimate based on the
1$-$0S(1) line flux is 1.12$\times10^2\msun$. This mass is an order of magnitude larger
than in the star-forming knots in the ring. A large amount of cold gas is also
found, which we estimated using the empirical relation derived by
\citet{Mazzalay13} for Seyfert galaxies and low-luminosity AGN. We derived a
cold gas mass of $\sim$8$\times10^7\msun$ for the nucleus of NGC\,613. These
values are larger than the hot and cold masses reported in \citet{Mazzalay13}
for a small sample of nearby AGN. Their sample includes two galaxies with
circumnuclear star-forming rings, NGC\,3351 and NGC\,4536, and the masses of hot
and cold gas measured by \citet{Mazzalay13} in those rings are much larger than
in their nuclei, which is opposite to what we find in NGC\,613. 

In contrast to the presence of hot and cold molecular gas, there is no
evidence for active star formation in the nucleus of NGC\,613: the recombination
lines of ionized hydrogen (e.g. \brg) are almost absent. This immediately raises
a number of questions about the mechanism responsible for heating the \htwo, the
amount of molecular gas within the nucleus, and the likely fate of this gas.

A similarity between the two galaxies with circumnuclear rings in
\citet{Mazzalay13} and NGC\,613 is the deficit of \brg\ emission relative to the
\htwo\ emission. However, the 1$-$0S(1)/\brg\ ratio that we measure in the
nucleus of NGC\,613 (14.9$\pm$3.0) is much larger than the values reported in
\citet{Mazzalay13} for NGC\,3351 and NGC\,4536 on the same scales (5.9 and 1.9).
This ratio is normally $\la$0.6 in starburst galaxies\footnote{In the ring hot
spots we measure 1$-$0S(1)/\brg$\leq$0.67.}, between 0.6 and 6 in Seyfert
galaxies, and larger than 6 in Low lonization Nuclear Emission Regions (LINERs;
\citealt{Riffel13,Mazzalay13}). The extreme value of 1$-$0S(1)/\brg\ measured in
the nucleus of NGC\,613 is thus consistent with a low-luminosity AGN (Seyfert or
LINER) with a strong influence from shock heating. The interaction with the
radio jet is thus likely enhancing the \htwo\ emission in the
nucleus.\looseness-2 

Using $K$-band {\tt SINFONI} data at the highest resolution available,
\citet{Davies07} studied the AGN--star formation connection in the inner 10\,pc
of nine nearby Seyferts. They found evidence for starbursts which took
place in the last 10--300\,Myr but are no longer active, and for a
delay of 50--100\,Myr between the triggering of the star formation in those
galaxies and the onset of the nuclear activity. Considering the lack of star
formation in the nucleus of NGC\,613, the large amount of cold molecular gas and
the confirmed presence of nuclear activity, we speculate with a cyclical
scenario, where a starburst episode was followed by another of nuclear activity.
In addition, the interaction with the radio jet likely produced a substantial
amount of molecular gas, creating a reservoir of cold gas that can sustain the
AGN and possibly fuel the next starburst episode.

We have attempted to date the last episode of star formation in the nucleus
comparing the \brg\ equivalent width with Starburst99 model
predictions. The measured value of 1.89\,\AA\ implies that the latest
starburst happened sometime between 8 and 13\,Myr ago (depending on the adopted
metallicity)\footnote{Considering the possible effect of the radio jet, this
estimate is necessarily a lower limit for the last starburst.}. Based on our
current estimate of H$_2$ molecular gas available, it is interesting to
determine when the next event might take place. The estimated free-fall time
(t$_{\rm ff}$), i.e., the earliest possible collapse time, for a cloud of
0\farcs5 radius (our aperture size) and $\sim$8$\times10^7\msun$ is t$_{\rm
ff}$\,$\sim$0.5\,Myr. The next starburst episode can thus be imminent
considering the dynamical timescales ($\sim$\,Gyr) in this type of galaxies.
Compared to the measurements presented by \citet{Mazzalay13} for a set of five
star-forming galaxies, the amount of H$_2$ present in NGC\,613 is at least a
factor of two, and up to a factor forty, larger for similar aperture
measurements. This comparison emphasises the rather special situation we
are observing in NGC\,613, where a starburst episode is likely about to
happen. This result, together with the lack of emission lines in other
galaxies in our original sample (paper I), suggest that as long as there is gas
supply towards the centre of the galaxy, recurrent star formation events can
take place in very short time intervals. This scenario is consistent with the
one established by \citet{Davies07}, although the time scales for NGC\,613 are
somewhat shorter. The lack of apparent gas transfer from the star-forming ring
to the inner regions, based on the H$_2$ emission-line map in
Fig.~\ref{fig:maps}, makes it rather difficult to determine the frequency of
these events in NGC\,613. While interesting, NGC\,613 is by no means a unique
case. Additional examples of galaxies, with large \htwo\ mass concentrations in
their nuclei, that might be at the verge of undergoing a starburst phase include
NGC\,6946 \citep{schinnerer06,schinnerer07} and NGC\,7552
\citep{pan13}.\looseness-2

While the discussion about recurrent star formation episodes presented here
focuses on the nucleus of NGC\,613, it is worth mentioning that a similar
behaviour has been observed in H{\sc ii} regions of star-forming nuclear rings 
and nuclear clusters (with intense starbursts followed by long inactive periods,
e.g.~\citealt{allard06,walcher06,sarzi07}).

% Simple aperture photometry on the H$_2$ map in
% Fig.~\ref{fig:maps} reveals that the full-width half maximum of the molecular
% gas present in the nucleus measures approximately 0\farcs78 ($\sim$66\,pc). This
% size is larger than the average seeing during our observations (0\farcs5), and
% thus the molecular cloud is resolved by our {\tt SINFONI} observations. We
% estimate the full extent (i.e. the diameter) of the cloud to be 1\farcs25
% ($\sim$106\,pc).

% G = 6.67259D-8  cm3 g-1 s-2
% Msun = 1.99D33 g
% 1 pc = 3.086D18 cm  
%
% Scale (pc/arcsec) = 84.8 pc/arcsec
% R = 0.5arcsec = 42.4 pc = 1.31D20 cm
% M = 8D7 Msun = 1.592D41 g
%
% density = (3*M)/(4*pi*R^3) = 1.69D-20 gr cm-3
% t_ff = sqrt((3*pi)/(32*G*density)) = 1.62D13 s = 512464 yr = 0.5 Myr

%###############################################################################
\section{SUMMARY AND CONCLUSIONS}\label{sec:summary}
%###############################################################################

In this paper we have made use of NIR integral-field {\tt SINFONI} observations to
study the inner regions of the nearby spiral galaxy NGC\,613. This galaxy is
part of the sample of five galaxies with star-forming nuclear rings presented in
\citet{boker08}. We selected this galaxy for a more detailed analysis due to the
peculiar nature of its circumnuclear environment: a star-forming ring, a radio
jet, and an unexpectedly high level of H$_2$ concentrated in its nucleus.

In \citet{boker08} we concluded that star formation along the inner ring
proceeds in a ''pearls on a string`` fashion, i.e. with star clusters in the
ring getting progressively older as they move away from the over-density
regions. The complexity of studying stellar content with the CO bandhead alone
prevented us from checking whether the same behaviour is observed in the
underlying populations. Nevertheless, the measurement of the CO equivalent width
(i.e. D$_{\rm CO}$ index), together with the $H-K$ colour, now allows us to
determine the amount of extinction in the innermost regions of NGC\,613.
Incidentally, the comparison of our D$_{\rm CO}$ values with those of early-type
galaxies reinforces previous findings in support of environmentally driven
carbon abundances.

The analysis of the emission lines reveals that the gas in the nucleus 
is not exclusively photoionized by the AGN, but also shock-heated. The hot 
spots along the ring however show a much more pronounced contribution from UV 
fluorescence. The star formation rates and masses of the ionised gas in 
H{\sc ii} regions are consistent with those observed in star-forming 
regions in other galaxies.

In our discussion, we used all this information to study the origin of the 
nuclear activity and establish the orientation of the observed radio outflow, 
which is nicely mapped by the \fetwo\ distribution and velocity dispersion. 
Based on the existence of a maser in the nucleus, we propose that the radio jet 
is likely oriented very close to the plane of the sky and that it is sweeping 
away the gas and dust in one portion of the ring.

Finally, we have investigated the fate of the unusually large amount of H$_2$
gas present in the nucleus of NGC\,613. We estimate that the last episode of
star formation took place around 10\,Myr ago and establish a lower
limit for the onset of the next event of 0.5\,Myr. The recurrence of
these episodes, however, is rather uncertain, especially since we have not found
any evidence for gas transfer from the star-forming ring to the nucleus. In
addition to that, one has to consider the role the radio jet may play in
inhibiting or fostering the generation of new stars. Still, datasets like the
one presented here have the important value of setting constraints and adding
fundamental information to the still rather poorly understood AGN--star
formation connection.\looseness-2

\section*{Acknowledgments}

The authors are indebted to J.~A. Acosta, A. Alonso-Herrero, P. Castangia, P.
Esquej and A. Vazdekis for insightful comments and suggestions at different
stages of this work. J.~F.-B. acknowledges support from the Ram\'on y Cajal
Program, grants AYA2010-21322-C03-02 from the Spanish Ministry of Economy and
Competitiveness (MINECO). We also acknowledge support from the FP7 Marie Curie
Actions of the European Commission, via the Initial Training Network DAGAL under
REA grant agreement number 289313. C.~R.~A. acknowledges grant PN
AYA2010-21887-C04.04 (Estallidos) from MINECO. Based on observations collected
at the European Southern Observatory, Chile, for proposal
076.B-0646(A).\looseness-2

\bibliographystyle{mn2e} % style mn2e.bst
\bibliography{sinfoni_NGC0613_astroph} % your references Yourfile.bib

%===============================================================================
% TABLES
%===============================================================================

%--------------------------------------------------------------------------------
\begin{table*}
\caption{Observed emission-line fluxes and derived properties}
\label{tab:fluxes}
\begin{center}
\begin{tabular}{cccccccccc}
\hline
\hline
Aperture & \fetwo & Br10 & \heone & \brg & A$_{\rm V}$ & Log($L_{{\rm H\alpha}}$) & SFR & Log(M$_{\rm HII}$) & D$_{\rm CO}$ \\
(1) & (2) & (3) & (4) & (5) & (6) & (7) & (8) & (9) & (10) \\
\hline
Nucleus & \phantom{0}85.50\,$\pm$\,3.28 &                     $\cdots$ &                     $\cdots$ & \phantom{0}4.82\,$\pm$\,0.89 & \phantom{0}3.62 & 39.26 & 0.02 & 3.10 & 1.207 \\
Spot~1  & \phantom{0}15.90\,$\pm$\,0.32 & \phantom{0}3.54\,$\pm$\,0.23 & \phantom{0}2.55\,$\pm$\,0.26 &           10.90\,$\pm$\,0.35 & \phantom{0}6.50 & 39.62 & 0.03 & 3.45 & 1.228 \\
Spot~2  & \phantom{0}25.70\,$\pm$\,0.39 & \phantom{0}4.78\,$\pm$\,0.23 & \phantom{0}3.81\,$\pm$\,0.27 &           14.70\,$\pm$\,0.32 & \phantom{0}7.24 & 39.75 & 0.04 & 3.58 & 1.230 \\
Spot~3  &           109.00\,$\pm$\,0.30 &           14.00\,$\pm$\,0.16 &           13.30\,$\pm$\,0.21 &           42.80\,$\pm$\,0.29 &           15.27 & 40.21 & 0.13 & 4.05 & 1.220 \\
Spot~4  & \phantom{0}53.00\,$\pm$\,0.35 &           11.30\,$\pm$\,0.22 & \phantom{0}9.51\,$\pm$\,0.29 &           34.60\,$\pm$\,0.37 & \phantom{0}9.40 & 40.12 & 0.10 & 3.95 & 1.213 \\
Spot~5  & \phantom{0}27.50\,$\pm$\,0.38 &           10.30\,$\pm$\,0.22 & \phantom{0}9.90\,$\pm$\,0.29 &           31.50\,$\pm$\,0.38 & \phantom{0}6.73 & 40.08 & 0.10 & 3.91 & 1.206 \\
Spot~6  & \phantom{0}29.90\,$\pm$\,0.32 & \phantom{0}8.22\,$\pm$\,0.19 & \phantom{0}5.81\,$\pm$\,0.21 &           25.20\,$\pm$\,0.31 & \phantom{0}9.43 & 39.98 & 0.08 & 3.82 & 1.149 \\
Spot~7  & \phantom{0}34.10\,$\pm$\,0.30 & \phantom{0}6.08\,$\pm$\,0.24 & \phantom{0}4.46\,$\pm$\,0.24 &           18.60\,$\pm$\,0.37 & \phantom{0}7.38 & 39.85 & 0.06 & 3.69 & 1.223 \\
Spot~8  & \phantom{00}4.82\,$\pm$\,0.60 &                     $\cdots$ &                     $\cdots$ & \phantom{0}1.49\,$\pm$\,0.37 & \phantom{0}0.84 & 38.75 & $\cdots$ & 2.59 & 1.219 \\
\hline
\hline
\end{tabular}
\end{center}
\begin{flushleft}
\small{NOTES: Fluxes and their uncertainties are corrected for internal extinction 
(based on the values in column 6) and are expressed in units of 10$^{-16}$ erg s$^{-1}$ cm$^{-2}$. 
$L_{{\rm H\alpha}}$ is in erg s$^{-1}$. SFR is expressed in $\msun$ yr$^{-1}$, and M$_{\rm HII}$ in $\msun$.}
\end{flushleft}
\end{table*}
%--------------------------------------------------------------------------------

%--------------------------------------------------------------------------------
\begin{table*}
\caption{\htwo\ emission-line fluxes and derived properties}
\label{tab:h2fluxes}
\begin{center}
\setlength{\tabcolsep}{3pt}
\begin{tabular}{cccccccccccc}
\hline
\hline
Aperture & 1$-$0~S(0) & 1$-$0~S(1) & 1$-$0~S(2) & 1$-$0~S(7) & 2$-$1~S(0) & 2$-$1~S(1) & 2$-$1~S(3) & 3$-$2~S(1) & 3$-$2~S(3) & $m_{\rm H_2}^{hot}$ & $m_{\rm H_2}^{cold}$\\
(1) & (2) & (3) & (4) & (5) & (6) & (7) & (8) & (9) & (10) & (11) & (12) \\
\hline
Nucleus &           20.70\,$\pm$\,0.92 &           71.90\,$\pm$\,1.24 &           25.00\,$\pm$\,0.94 &           16.40\,$\pm$\,0.84 &          $\cdots$ &           10.90\,$\pm$\,0.94 & 5.04\,$\pm$\,0.80 &          $\cdots$ &          $\cdots$ &           112.0 &           80.6\\
Spot~1  & \phantom{0}2.67\,$\pm$\,0.29 & \phantom{0}6.05\,$\pm$\,0.35 & \phantom{0}2.34\,$\pm$\,0.25 &                     $\cdots$ &          $\cdots$ & \phantom{0}2.22\,$\pm$\,0.28 &          $\cdots$ &          $\cdots$ &          $\cdots$ & \phantom{00}9.4 & \phantom{0}6.8\\
Spot~2  & \phantom{0}3.46\,$\pm$\,0.33 & \phantom{0}7.89\,$\pm$\,0.44 & \phantom{0}3.31\,$\pm$\,0.33 &                     $\cdots$ &          $\cdots$ & \phantom{0}3.15\,$\pm$\,0.35 &          $\cdots$ &          $\cdots$ &          $\cdots$ & \phantom{0}12.3 & \phantom{0}8.8\\
Spot~3  & \phantom{0}8.88\,$\pm$\,0.21 &           24.70\,$\pm$\,0.28 & \phantom{0}9.03\,$\pm$\,0.22 &           11.70\,$\pm$\,0.19 & 3.59\,$\pm$\,0.22 & \phantom{0}6.57\,$\pm$\,0.23 &          $\cdots$ &          $\cdots$ & 3.11\,$\pm$\,0.20 & \phantom{0}38.5 &           27.7\\
Spot~4  & \phantom{0}5.23\,$\pm$\,0.28 &           13.80\,$\pm$\,0.36 & \phantom{0}5.34\,$\pm$\,0.31 &                     $\cdots$ & 2.28\,$\pm$\,0.29 & \phantom{0}4.20\,$\pm$\,0.31 &          $\cdots$ &          $\cdots$ &          $\cdots$ & \phantom{0}21.5 &           15.5\\
Spot~5  & \phantom{0}2.33\,$\pm$\,0.27 & \phantom{0}6.39\,$\pm$\,0.37 &                     $\cdots$ &                     $\cdots$ &          $\cdots$ & \phantom{0}2.05\,$\pm$\,0.32 &          $\cdots$ &          $\cdots$ & 1.92\,$\pm$\,0.27 & \phantom{0}10.0 & \phantom{0}7.2\\
Spot~6  & \phantom{0}3.21\,$\pm$\,0.25 &           11.70\,$\pm$\,0.34 & \phantom{0}4.02\,$\pm$\,0.24 & \phantom{0}2.81\,$\pm$\,0.22 &          $\cdots$ & \phantom{0}2.57\,$\pm$\,0.27 &          $\cdots$ & 2.14\,$\pm$\,0.26 &          $\cdots$ & \phantom{0}18.2 &           13.1\\
Spot~7  & \phantom{0}4.67\,$\pm$\,0.29 &           12.60\,$\pm$\,0.36 & \phantom{0}5.16\,$\pm$\,0.27 & \phantom{0}2.56\,$\pm$\,0.23 &          $\cdots$ & \phantom{0}3.23\,$\pm$\,0.29 &          $\cdots$ & 1.71\,$\pm$\,0.28 &          $\cdots$ & \phantom{0}19.6 &           14.1\\
Spot~8  &                     $\cdots$ & \phantom{0}2.11\,$\pm$\,0.48 & \phantom{0}1.18\,$\pm$\,0.35 &                     $\cdots$ &          $\cdots$ &                     $\cdots$ &          $\cdots$ &          $\cdots$ &          $\cdots$ & \phantom{00}3.3 & \phantom{0}2.4\\
\hline
\hline
\end{tabular}
\end{center}
\begin{flushleft}
\small{NOTES: Fluxes and their uncertainties are corrected for internal extinction 
(based on the values in column 6 of Table~\ref{tab:fluxes}) and are expressed in 
units of 10$^{-16}$ erg s$^{-1}$ cm$^{-2}$. \htwo\ (hot and cold) masses are based on 
calibrations from \citet{Reunanen02} and \citet{Mazzalay13}, and 
are expressed in units of $\msun$ and 10$^{6}\msun$, respectively.}\end{flushleft}
\end{table*}
%--------------------------------------------------------------------------------
 
\label{lastpage}

\end{document}